%Paper: hep-ph/9409373
%From: rball@surya11.cern.ch (Richard Ball)
%Date: Wed, 21 Sep 94 17:06:19 +0200

%%%%%%%%%%%%%%%%%%%%%%%%%%%%%%% v3.tex %%%%%%%%%%%%%%%%%%%%%%%%%%%%%%%%%%
%%%%%%%               The Rise in F_2^p at HERA                    %%%%%%
%%%%%%%                 R. D. Ball and S. Forte                    %%%%%%
%%%%%%%      in plain TeX with the harvmac macro package           %%%%%%
%%%%%%%    6 figures in compressed postscript in separate file     %%%%%%
%%%%%%%      to be printed separately (as 3 postscript files)      %%%%%%
%%%%%%%%%%%%%%%%%%%%%%%%%%%%%%%%%%%%%%%%%%%%%%%%%%%%%%%%%%%%%%%%%%%%%%%%%

\input harvmac

%\draftmode
\noblackbox
\pageno=0\nopagenumbers\tolerance=10000\hfuzz=5pt
\line{\hfill CERN-TH.7421/94}
\vskip 36pt
\centerline{\bf THE RISE IN $F_2^p$ AT HERA}
\vskip 36pt\centerline{Richard~D.~Ball\footnote{$^\dagger$}{On leave
from a Royal Society University Research Fellowship.}
 and Stefano~Forte\footnote{$^\ddagger$}{On leave
from INFN, Sezione di Torino, Italy.}}
\vskip 12pt
\centerline{\it Theory Division, CERN,}
\centerline{\it CH-1211 Gen\`eve 23, Switzerland.}
\vskip 60pt
{\medskip\narrower
\ninepoint\baselineskip=9pt plus 2pt minus 1pt
\lineskiplimit=1pt \lineskip=2pt
\centerline{\bf Abstract}
\noindent
We show that the rise in $F_2^p$ at small $x$ and large $Q^2$ seen at HERA
is indeed the non-Regge double asymptotic scaling behaviour expected
from the perturbative emission of strongly ordered hard gluons. An
alternative explanation, in which there is no strong ordering, and a
new hard Reggeon is generated,
is also tried but found wanting: its theoretical short-comings
are betrayed by
its failure to properly account for the HERA data.
}
\vfill
\centerline{Talk given at ``The Heart of the Matter'', Blois,
                          June 1994,}
\centerline{to be published in the proceedings (Editions Fronti\`eres).}
\vskip 20pt
\line{CERN-TH.7421/94\hfill}
\line{September 1994\hfill}

\vfill\eject
\footline={\hss\tenrm\folio\hss}
%\nopagenumbers

%#############################################################################%

\def\frac#1#2{{{#1}\over {#2}}}
\def\half{\hbox{${1\over 2}$}}

\def\smallfrac#1#2{\hbox{${{#1}\over {#2}}$}}

\def\GeV{{\rm GeV}}

\catcode`@=11 %This allows us to modify plain macros
\def\slash#1{\mathord{\mathpalette\c@ncel#1}}
 \def\c@ncel#1#2{\ooalign{$\hfil#1\mkern1mu/\hfil$\crcr$#1#2$}}
\def\lsim{\mathrel{\mathpalette\@versim<}}
\def\gsim{\mathrel{\mathpalette\@versim>}}
 \def\@versim#1#2{\lower0.2ex\vbox{\baselineskip\z@skip\lineskip\z@skip
       \lineskiplimit\z@\ialign{$\m@th#1\hfil##$\crcr#2\crcr\sim\crcr}}}
\catcode`@=12 %at signs are no longer letters

\def\DZP{\hbox{D0$'$}}\def\DMP{\hbox{D-$'$}}

\def\PR{{\it Phys.~Rev.~}}

\def\NP{{\it Nucl.~Phys.~}}
\def\NPBPS{{\it Nucl.~Phys.~B (Proc.~Suppl.)~}}
\def\PL{{\it Phys.~Lett.~}}
\def\PRep{{\it Phys.~Rep.~}}

\def\ZP{{\it Zeit.~Phys.~}}

\def\vol#1{{\bf #1}}\def\vyp#1#2#3{\vol{#1} (#2) #3}

%$$$$$$$$$$$$$$$$$$$$$$$$$$$$$$$$$$$$$$$$$$$$$$$$$$$$$$$$$$$$$$$$$$$$$$$$$$$$%

\nref\HERA{ M.~Roco \& K.~M\"uller, results from
                  ZEUS \& H1, 29th Rencontre de Moriond, March 1994.}
\nref\DAS{ R.D.~Ball \& S.~Forte, \PL\vyp{B335}{1994}{77}.}
       %   CERN-TH.7265/94, {\tt hep-ph/9405320}.
\nref\Test{ R.D.~Ball \& S.~Forte, CERN-TH.7331/94,
            \PL{\bf B} (in press).}
       % {\tt hep-ph/9406385}.
\nref\DGPTWZ{ A.~De~Rujula et al, \PR\vyp{D10}{1974}{1649}.}
\nref\Mont{ R.D.~Ball \& S.~Forte, CERN-TH.7422/94,
             \NPBPS\ (in press).}
       % {\tt hep-ph/940????}.
\nref\MRS{ A.D.~Martin et al, \PL\vyp{B306}{1993}{145}.}
\nref\HOS{ T.~Jaroszewicz, \PL\vyp{B116}{1982}{291}\semi
           S.~Catani \& F.~Hautmann, \PL\vyp{B315}{1993}{157},
                         Cavendish-HEP-94-01.}
\nref\EKL{ R.K.~Ellis et al, \NP\vyp{B420}{1994}{517}.}
\nref\Screen{J.~Kwiecinski,~\ZP\vyp{29}{1985}{147}\semi
                 A.H.~Mueller \& J.~Qiu, \NP\vyp{B268}{1986}{427}.}
\nref\Cat{ S.~Catani, DFF~207/6/94.}
\nref\Keqn{ L.V.~Gribov et al, \PRep\vyp{100}{1983}{1}\semi
                J.~Kwiecinski, \ZP\vyp{29}{1985}{561}.}
\nref\MWeqn{ G.~Marchesini \& B.R.~Webber,
                   \NP\vyp{B349}{1991}{617}\semi
                E.M.~Levin et al, \NP\vyp{B357}{1991}{167}.}
\nref\runLip{C.~Lovelace, \NP\vyp{B95}{1975}{12}\semi
              J.C.~Collins \& J.~Kwiecinski, \NP\vyp{B316}{1989}{307}.}
\nref\ColEll{ J.C.~Collins \& R.K.~Ellis,
                  \NPBPS\vyp{18C}{1990}{80}; \NP\vyp{B360}{1991}{3}.}
\nref\ColLan{ J.C.~Collins \& P.V.~Landshoff,
                      \PL\vyp{B276}{1992}{196}.}
\nref\HardScr{ E.~Laenen et al, \NP\vyp{B419}{1994}{39}.}
\nref\MRSH{ A.D.~Martin et al, RAL-94-055, DTP/94/34.}

\nfig\rise{The rise in $F_2^p$: $R_F'F_2^p$ vs. $\sigma$. Also shown
is the double scaling prediction\DAS, and a fitted straight line for
comparison.}
\nfig\kin{The $x$-$t$ plane, showing the different kinematic
regions, and in particular the double scaling region, with scaling
co-ordinates $\sigma$ and $\rho$.}
\nfig\sp1{Double scaling plots of $R_F F_2^p$ vs. a) $\sigma$ and b)
$\rho$. The data are taken from ref.\HERA, and the curves are those of
the MRS parton distributions \DZP. The double scaling curves \DAS\ are
shown dotted.}
\nfig\sp2{As fig.~3, but with the curves now corresponding to \DMP\
and the `hard pomeron' curves computed in ref.\DAS (dotted).}

The structure function $F_2^p(x,Q^2)$, recently measured\HERA\ by the HERA
experiments ZEUS and H1 in the hitherto unexplored
region $10^{-4}\lsim x\lsim 10^{-2}$, $5\GeV^2\lsim Q^2 \lsim 10^5 \GeV^2$,
rises dramatically both as $x$ decreases and $Q^2$ increases. The form
of this rise may be clearly exhibited by using the variables\DAS
\eqn\sr{
\sigma\equiv\sqrt{\ln\smallfrac{x_0}{x}\ln\smallfrac{t}{t_0}},
\qquad\rho\equiv\sqrt{\ln\smallfrac{x_0}{x}\big/\ln\smallfrac{t}{t_0}},}
where $t\equiv\ln(Q^2/\Lambda^2)$. Rescaling $F_2^p$ by a
simple multiplicative factor $R_F'\equiv N\sigma^{1/2}\rho
e^{\delta\sigma/\rho}$,
where $\delta =\smallfrac{61}{45}$, we may then plot it on a
logarithmic scale against $\sigma$. The resulting
plot (fig.~1) is interesting in two respects: firstly when $\sigma$ is large
enough all the data lie on a single line, quite independently of the value of
$\rho$ (provided $\rho$ too is large enough),
and secondly the rise of $\log R_F'F_2^p$ with $\sigma$ is
linear.\DAS\ The slope of the rise\Test\ is $2.37\pm 0.16$ (dotted
line in fig.~1).

%\vskip 8truecm
%\topinsert\vbox{\vskip 60mm}\endinsert

Remarkable though it is, such a rise was not unanticipated: the behaviour of
$F_2$ at small $x$ was computed in perturbative QCD soon after the discovery
of asymptotic freedom\DGPTWZ, using the operator product expansion
(at leading twist), the renormalization group, perturbation theory (at one
loop) and assuming that at some low starting scale $Q_0$ the small $x$
behaviour of $F_2$ is given by conventional Regge behaviour, and thus
flat. The resulting perturbatively generated non-Regge behaviour takes
the form
\eqn\Fsoft{
F_2\sim {\cal N}f(\smallfrac{\gamma}{\rho})(R_F')^{-1}e^{2\gamma\sigma},}
where ${\cal N}$ and $f\sim 1 + O(\rho^{-1})$ depend on the
(soft) non-perturbative input at $Q_0$.
Since $2\gamma\equiv 4\sqrt{N_c/\beta_0} = 2.4$ is simply a
numerical constant, perturbative QCD thus predicts precisely the
double scaling behaviour\DAS\ described above: when $F_2$ is
rescaled by a factor
$R_F\equiv e^{-2\gamma\sigma}R_F'$ to remove both the linear rise and
the sub-asymptotic effects $R_F'$, the
resulting structure function is asymptotically independent of
both $\sigma$ and $\rho$.

The rise in $F_2$ is thus generated perturbatively through iteration of the
simple processes \hbox{$g\to gg$} and \hbox{$g\to \bar{q}q$:} the
splitting functions
$P_{gg}(x)$ and $P_{qg}(x)$ are both singular at leading order
as \hbox{$x\to 0$.}
In fact at small $x$ the gluon evolution equation reduces to a
wave equation in light-cone variables $\xi\equiv\log\smallfrac{x_0}{x}$,
$\zeta\equiv\log\smallfrac{t}{t_0}$, with a (negative)
mass term which generates an exponential growth in $xg$ inside the light-cone
(see fig.~2: turn it anti-clockwise through $45^o$). This then drives a
similar growth in $F_2$. This explains why the natural variables
to use when $x<x_0$ and $t>t_0$ are $\sigma = \sqrt{\xi\zeta}$ and
$\rho = \sqrt{\xi/\zeta}$, and also why a generic exponential rise
($\sigma$-scaling) will be produced isotropically ($\rho$-scaling) whenever
the boundary conditions set at $x=x_0$ and $t=t_0$ are sufficiently soft
(and in particular $g(x;t_0)\sim x^{-1}$ as $x\to 0$), so
that the main source of the waves is close to the origin.

%\vskip 8truecm
%\topinsert\vbox{\vskip 60mm}\endinsert

Double asymptotic scaling seems to set in just as precociously as Bjorken
scaling: almost all the HERA data falls within the scaling region.
For this reason we take $Q_0= 1~\GeV$; higher
values would imply large double scaling violations not seen in the data.
To illustrate this point we display in fig.~3 the MRS prediction\MRS\ \DZP,
a soft distribution fitted to pre-HERA data,
and then evolved (at two loops) from $Q_0^2 = 4~\GeV^2$. Although
similar in shape to the double scaling
curves\DAS\ (shown dotted in fig.~3), \DZP\ agrees less well with
the HERA data.
Dropping the starting scale to $1~\GeV$ would have given a prediction which,
while having the same shape as double scaling, also had the correct
normalization.\Mont

Note that although $1~\GeV$ is at the boundary of the nonperturbative region,
we only compare to data with $Q^2\gsim 5\GeV^2$, where higher twist
corrections to $F_2$ are negligible. Higher twist evolution below this
scale may be absorbed into similar uncertainties in the starting distribution.

While double scaling captures the essential physics of the
rise in $F_2$, there are several corrections still to be included:

\noindent
i) Sub-asymptotic corrections, both at one and two loops;
these may (and have been\DAS\Mont) computed explicitly; the dominant
uncertainty is now the shape of the initial gluon distribution.

\noindent
ii) Post-asymptotic corrections, due to the leading singularities in the
anomalous dimensions at higher orders in $\alpha_s$; these are now known
explicitly\HOS\ and may be included in the Altarelli-Parisi equation to
all orders. In this way the the strong $k_T$
ordering implicit at leading order is gradually relaxed in a controlled way
as we go to smaller $x$. In the HERA region such corrections are still quite
small however\EKL.\footnote{*}{Shift the axes of fig.~9 of
ref.\EKL\
by $\smallfrac{1}{10}$ and $\smallfrac{1}{4}$ respectively, to take into
account our choice of $x_0$ and $Q_0$: almost all the HERA data then lies
within these authors' convergence criterion.} Subleading singularities
might also be calculable.

\noindent
iii) Higher twist (screening) corrections, necessary to restore unitarity
post-asymptotically; these are probably very small in the HERA
region\Screen, though they are difficult to compute reliably.

Thus in the `soft pomeron' approach (perturbative evolution of
an initially soft nonperturbative input), higher order corrections are all
(except possibly iii) under control, in the sense that they are small
in the HERA region, and may furthermore be
systematically computed within a well defined scheme. This is essentially
because the operator product expansion and renormalization group are used to
control the expansions.
By contrast in the `hard pomeron' approach (see
ref.\Cat\ for a critical review) these fundamentals
are abandoned in favour of a particular diagrammatic resummation.
This inevitably leads to difficulties.

The starting point for this latter approach is the `BFKL equation'
which resums logs
of $1/x$ at fixed $Q^2$; $k_T$ ordering is abandoned at the outset.
Solution of the equation gives
asymptotically (at small $x$ and large $Q^2$) the Regge-like behaviour
\eqn\Fhard{
F_2\sim {\cal N}\sqrt{Q^2/Q_0^2}\;x^{-\lambda},\qquad
\lambda = 12\log 2 \frac{\alpha_s}{\pi}.}
Such a rise violates double scaling\DAS, and indeed is
much stronger than that seen at HERA (it would imply an exponential rather
than linear rise in fig.~1, for example). However it need not be taken
too literally, because a number of important theoretical problems remain
unresolved:

\noindent
I) The BFKL equation does not consider evolution in $Q^2$, and thus does not
resum leading logs. The equation is thus inconsistent with the
renormalization group, and its solution cannot be related to known
distributions at larger $x$. Alternative equations\Keqn\MWeqn,
which also
have no $k_T$ ordering but include $Q^2$ evolution, are difficult to solve,
but seem to give numerical results consistent with leading order
Altarelli-Parisi evolution. This is because
when the coupling runs, the hard pomeron cut dissolves into a sequence
of poles\runLip, and the right-most pole has only a small residue.

\noindent
II) When the $k_T$ ordering implicit in the conventional leading twist
evolution equations is abandoned, contributions with different twist become
mixed together. This leads to nonperturbative infrared effects which
cannot be easily factored out, except by projecting onto the leading
twist.\HOS An attempt to solve this problem may be found in
ref.\ColEll. The implementation of the (kinematically necessary)
ultraviolet cutoff on $k_T$ is also problematic.\ColLan
Subleading corrections, while probably important (to ensure
momentum conservation, for example), are as yet unknown.

\noindent
III) Since unitarity is now violated more quickly since the growth (4)
is much stronger than (2), screening corrections are now much
larger\HardScr; they are also even more difficult to
calculate.

While most of these problems remain unresolved, it is not yet really possible
to compare the `hard' approach with the data. Despite (or perhaps
because of) this, it has recently become popular to employ a `hybrid'
approach, in which the perturbatively based small-$x$  behaviour (3) is used
as a nonperturbative boundary condition at $Q_0$, which is then evolved
using the conventional two loop evolution equations. The resulting $Q^2$
dependence is then relatively weak, as the Lipatov cut is to the right of
the leading singularities in the anomalous dimensions.\Cat\ Since such an
approach makes little sense theoretically, it is rather satisfying that the
data themselves contradict it: a hard starting distribution propagates inside
the light-cone anisotropically, generating strong violations of
$\rho$-scaling.\DAS This is demonstrated rather clearly
by the MRS prediction\MRS\ \DMP, displayed in fig.~4 (the `hard'
pomeron curves of ref.\DAS, also with $\lambda=\half$, are shown
dotted for comparison). In fact
\DMP\ fits the data even less well than \DZP, essentially because the
sharp rise at small $x$ is not matched by a corresponding rise at
large $Q^2$. Of course by reducing $\lambda$ and/or using
a boundary condition with only a very small relative admixture
of hard to soft behaviour, it is still possible to fit the data:
the recent MRS H fit mimics double scaling in precisely this way.
\footnote{*}{In fact there is only around $3\%$ of the hard solution in MRS(H):
see eqn(24) of ref.\MRSH.}

%\topinsert\vbox{\vskip 80mm}\endinsert
%\eject\vskip 8truecm\noindent

In conclusion, the `hard' pomeron is neither a proven consequence of
perturbative QCD, nor is it as yet visible in the structure function
data. Indeed, the simple prediction of perturbative QCD\DGPTWZ\ has
been beautifully confirmed\DAS\Test\ by the measured\HERA\ rise
in $F_2$ at HERA. This is fortunate, because it means that
perturbative QCD can be used to make reliable predictions at small $x$. It
should thus be possible to use the HERA data to further test QCD, measure
$\alpha_s$, and predict QCD backgrounds at the LHC by extrapolating to
yet higher values of $\sigma$.

\bigskip\noindent
{{\bf Acknowledgements:} We would like to thank
S.~Catani, R.K.~Ellis, F.~Hautmann, Z.~Kunszt, J.~Kwiecinski,
P.V.~Landshoff, E.M.~Levin, R.G.~Roberts and D.A.~Ross for discussions.
}

%$$$$$$$$$$$$$$$$$$$$$$$$$$$$$$$$$$$$$$$$$$$$$$$$$$$$$$$$$$$$$$$$$$$$$$$$$$$$$%

\bigskip
\listrefs
\listfigs
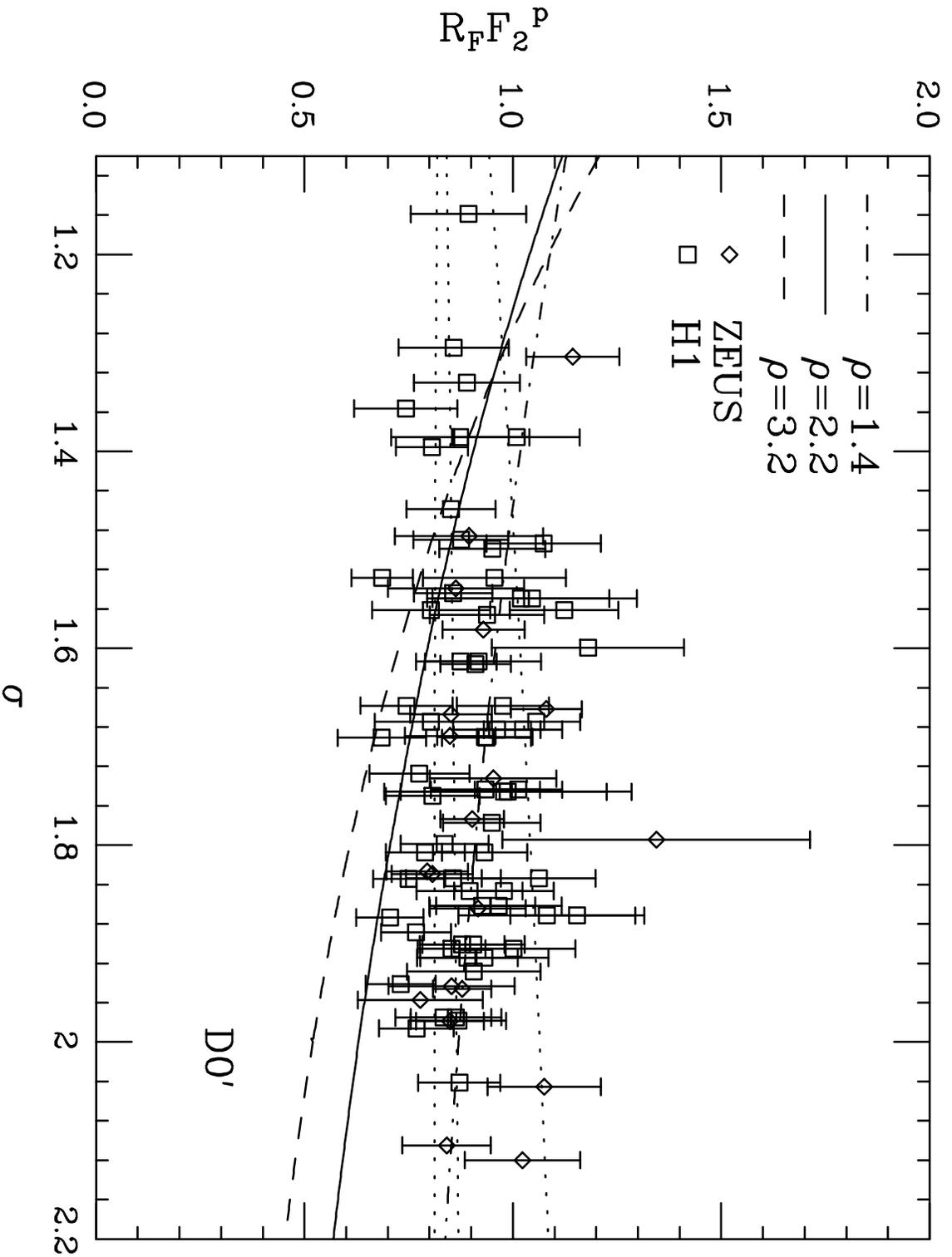
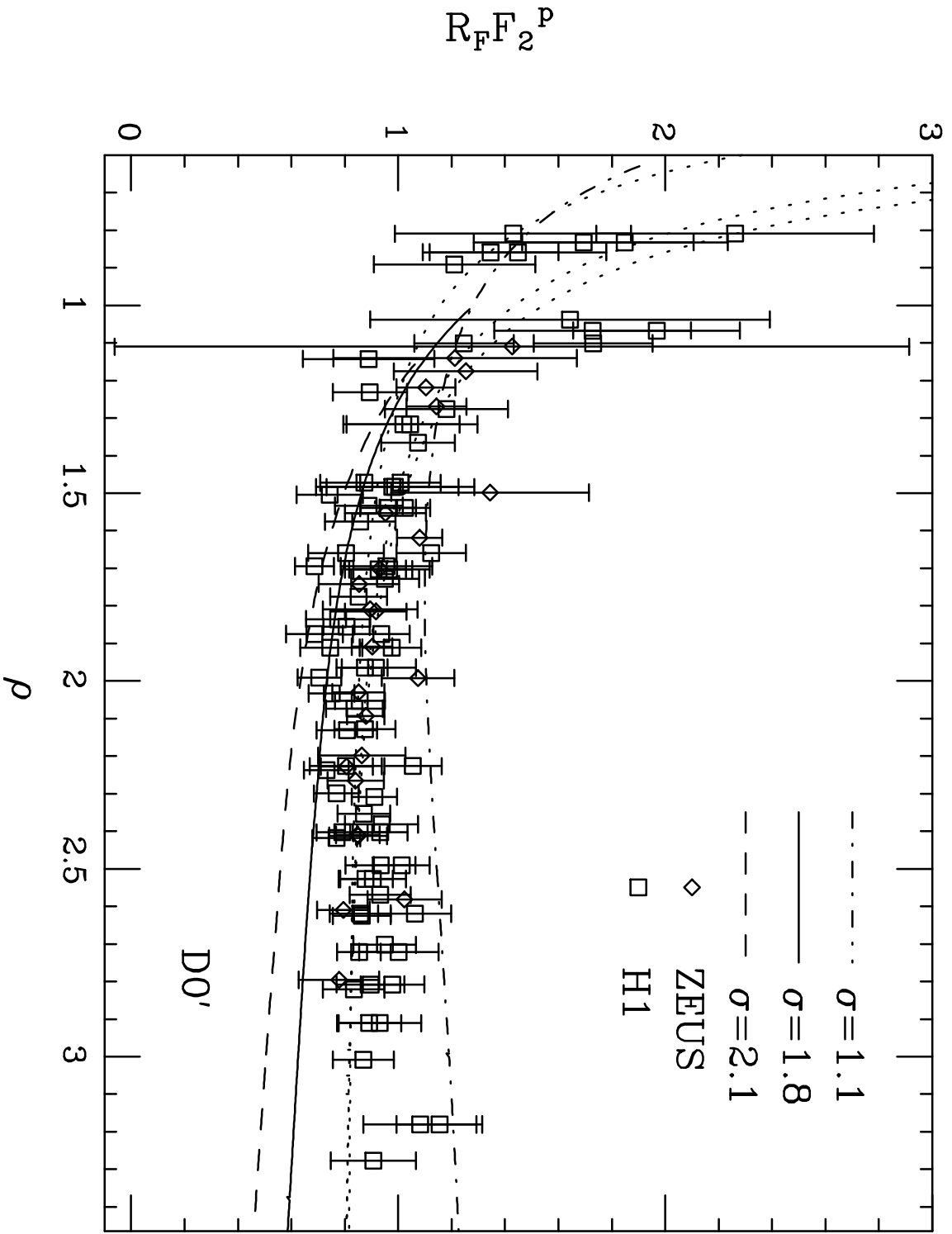
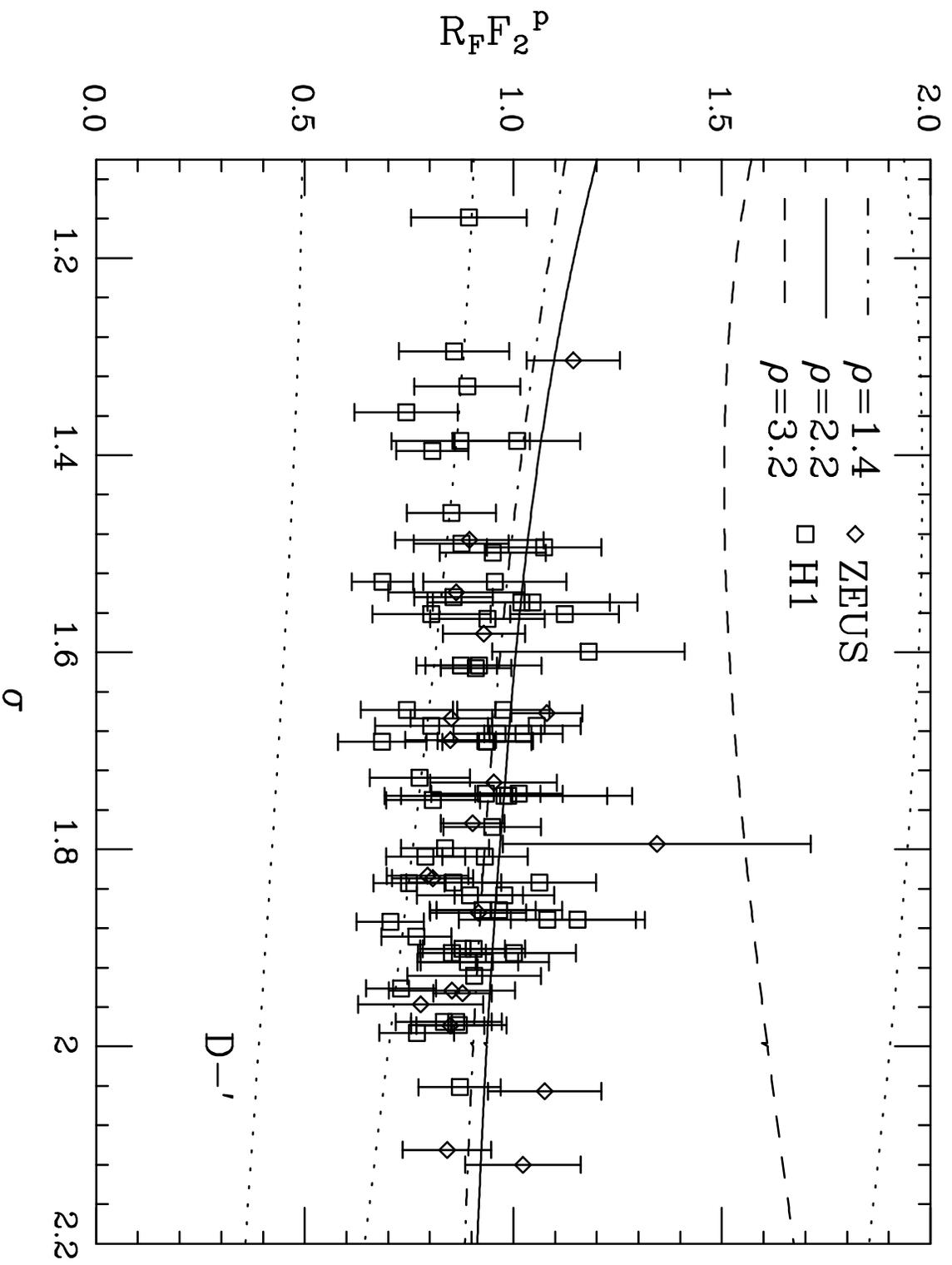
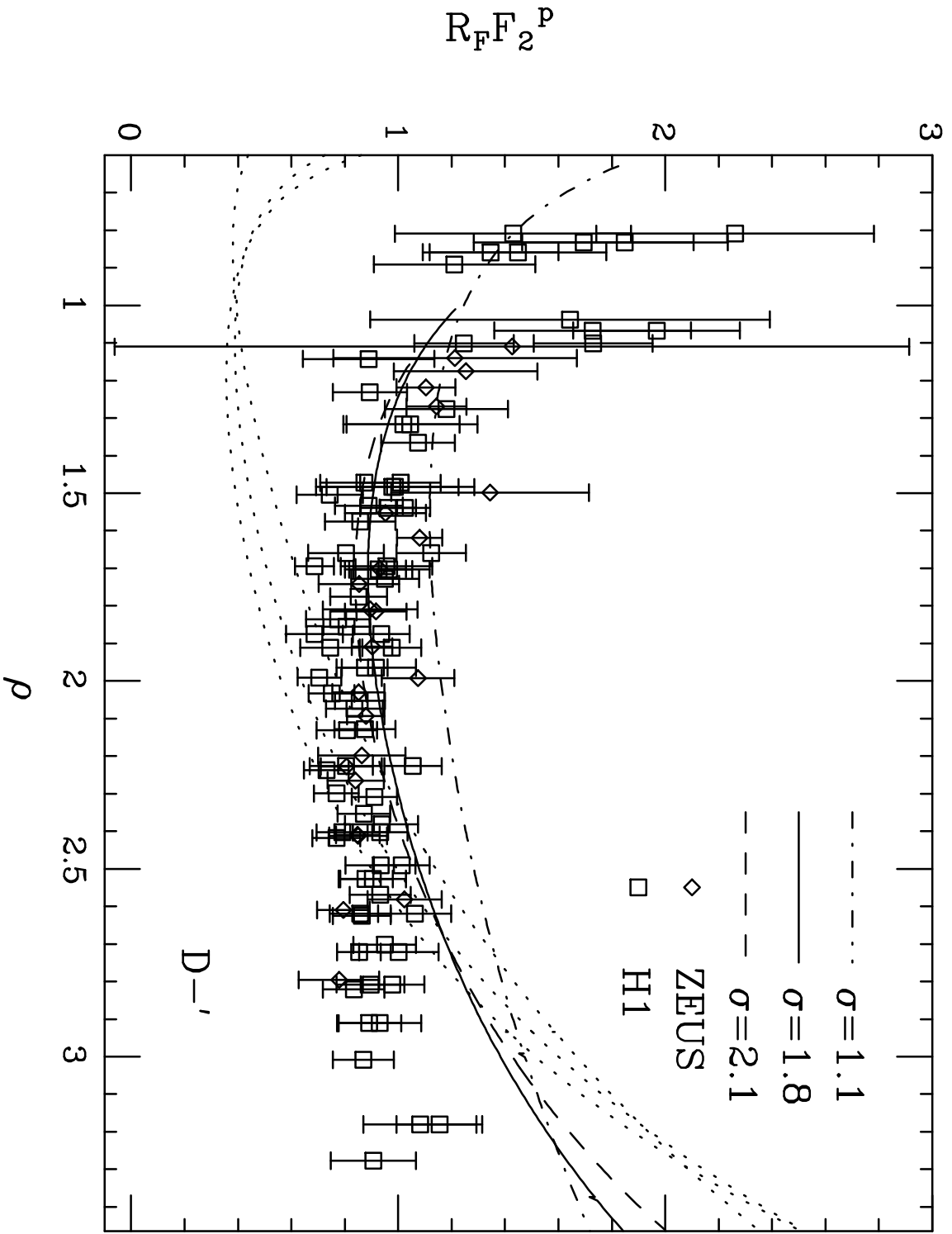
\end